\documentclass[12pt]{article}

\newcommand{\p}{\partial}

\begin{document}
\topmargin 0pt
\oddsidemargin 0mm

\renewcommand{\thefootnote}{\fnsymbol{footnote}}
\begin{titlepage}
\begin{flushright}UUPHY/03/7\\

math-ph/0306069
\end{flushright}

\vspace{5mm}
\begin{center}
{\Large \bf Symmetry analysis for a charged particle in a certain varying
magnetic field }
\vspace{10mm}

{\large
Karmadeva Maharana$^{*, \dag}$ }\\
\vspace{4mm}

\footnote{Present address}{\em  Department of Mathematics,  \\
Massachusetts Institute of Technology, Cambridge MA 02139, USA  \\
  email : maharana@math.mit.edu  \\
and \\
\footnote{Permanent address}Physics Department, Utkal University, Bhubaneswar 
751 004, India  \\ 
  email: karmadev@iopb.res.in }

\end{center}
\vspace{5mm}
\centerline{{\bf{Abstract}}}
\vspace{5mm}

We  analyze  the classical equations of motion for a
particle moving in the presence of a static  magnetic field 
applied in the $ z $ direction, which varies as $ {1\over{x^2}} $. 
We find the symmetries through Lie's
method of group analysis. In the corresponding quantum mechanical case,
the  method of spectrum generating $su(1,1)$ algebra is used to
find the energy levels for the  Schroedinger
equation without explicitly solving the equation. The
Lie point
symmetries are enumerated. We also find that for specific eigenvalues
the vector field contains $ {1\over{ x}} {{\p}\over{\p x}}$ and 
$ {1\over {x^2}} {{\p}\over{\p {x}}}$ type of terms and a finite
Lie product of the generators do not close.

 
\end{titlepage}

\newpage

\section{Introduction}


   
   We use some techniques from group theory to analyse the model physical 
   situation of
   a charged particle in  magnetic field $ B_x = 0 , B_y = 0 $ , and
   $ B_z = {1\over{x^2}} $ . Usually the symmetries of such systems are
   found with the use of  Noether's theorem by setting up a Lagrangian.
   However, a Lagrangian formulation of the problem sometimes 
   becomes difficult,
   as in Witten's example {\cite{witten}} of a classical Wess Zumino model{\cite{wess},
   representing the motion of a charged particle in the presence of a
   magnetic monopole and constrained to move on the surface of the sphere.
   The equations of motion in that case cannot be obtained from a Lagrangian
   as no Lagrangian can be written whose variation will give the desired
   equation of motion, if we restrict ourselves not going to a higher
   dimensional theory.  
   So in this case the usual Noether's theorem is difficult to apply and the
   generalised method has to be followed  to
   find the symmetries from the classical equation of motion 
      {\cite{stephani},\cite{olver}
  . Finding the 
   generators of the symmetries will inform us which symmetries exist and
   which of the symmetries are broken when the magnetic field is applied. 
     
     We also consider the related quantum mechanical case. The method 
       of spectrum
     generating algebra is applied to obtain the energy eigenvalues
     without solving the eigenvalue equation. 
     
     The quantum case is dealt first in a manner analogous to 
     the treatment followed by Landau to explain the 
     diamagnetism arising out of
     conduction electrons
     . The Schr\"odinger equation separates to a second order differential
     equation involving $x$. The energy eigenvalues for this equation
     can be obtained by mapping this problem to that of the spectrum
     generating algebra of $ SU (1,1) $ by identifying the generators 
     of the algebra with the {(differential )} operator realization
     of the algebra, and calculating the Casimir invariant.
     
     We  try to find
     the Lie point symmetries of this differential equation  
     and it is found that, when certain relations are satisfied 
      by the coefficients  of the different terms,some symmetry exist.
      The  vector fields representing the symmetries do not close
     under the Lie product. The fact that the Schr\"odinger equation
     can have solutions when certain symmetries exist is verified in
     three cases.

   \section{ The quantum eigenvalues}
   
   The Schr\"odinger equation for a spinless charged particle
 in the presence of a magnetic field is obtained by the usual 
 procedure of replacing
$p^a$ by $ p^a + {e\over c } A^a $ in the free particle equation
\cite{schiffhaug}. For the time being we consider any electrostatic
 potential to be absent. Here $A^a $ is the vector potential and we
 choose a particular gauge where
it is given by $ {\bf A} = \{ A_x = 0 , A_y = {{\cal B}\over x}  ,
A_z = 0 \} $ so that the magnetic
 field $ {\bf B } = {\bf \nabla} \times {\bf A} $
is $ \{ B_x = 0 , B_y = 0 , B_z = {-} {{\cal B}\over {x^2}} \} $. Here
$ \cal B $ is a constant. Thus the Hamiltonian operator becomes
\begin{eqnarray}
 H = {{1\over 2m } {( {\bf p } + {e\over c} {\bf A} )}^2 }
   =  {{ -} {{{{\hbar}^2}\over 2m}}  {\nabla}^2}
   + {{\hbar}\over{i}} { e\over mc}
{\cal B}  {1\over x} {\p\over\p{y}}  + {{e^2 {\cal B}^2 }\over{2mc^2}}
{1\over x^2 } .
\end{eqnarray}
Denoting 
\begin{eqnarray}
 {{e \cal B }\over{2mc}} = {\omega}_{\cal B}
\end{eqnarray}
the Schr\"odinger equation becomes
\begin{eqnarray}
 [ - {{{\hbar}^2}\over 2m } {\nabla}^2 
 + 2 {{\hbar}\over{i} } {\omega}_{\cal B}
  {1\over x} {\p\over\p{y}}  + 2m {{\omega}_{\cal B}}^2
{1\over x^2 } ] \phi  = E \phi  .
\end{eqnarray}
The second and third terms have only explicit $x$ dependence and we may use
\begin{eqnarray}
{\phi}(x, y, z ) = u(x) e^{ i {( k_y y + k_z z )}}
\end{eqnarray}
to separate out the $x$ dependent part. We thus obtain, for $u(x)$, the
equation
\begin{eqnarray}
 - {{{\hbar}^2}\over 2m} {{ d^2 u}\over{{dx}^2}} +
{1\over 2m} {({ \hbar}{ k_y } + {{2m {\omega}_{\cal B}}\over{x}}  )}^2 {u}
= {( E - {{{\hbar}^2 {k_z}^2 }\over 2m} )} u  \label{eq:SS}  .
\end{eqnarray} 
It is interesting to note that the Scr\"odinger equation for Kratzer's
molecular potential  is a special case of the above
equation.
The Kratzer molecular potential is of the form , in our notation,
\begin{eqnarray}
V (x) \propto  ( { a\over x } - {1\over 2} {{a^2}\over{x^2 }} )
\end{eqnarray}
and the eigenvalue equation is solved by the usual series expansion method.
The eigenfunctions turn out to be a general type of Kummer's hypergeometric
function $ _1F_1 $\cite{flugg}.

 Also  other (electrostatic) potentials of the form
$ 1\over x $ and $ 1\over{x^2} $ can be included in the above equation,
that will just change the appropriate coefficients C, D and E.
 Cordero and Ghirardi have calculated the energy eigenvalues
 for such cases using the
spectrum generating algebra method \cite{cordero}.
We write the equation (\ref{eq:SS}) as
\begin{eqnarray}
{{ d^2 u}\over{{dx}^2}} + {( {C\over{x^2}}  + {D\over{x}} + {\hat E} )} u 
= 0  \label{eq:BE}
\end{eqnarray}
where
\begin{eqnarray}
C = {-} {{4 m^2 {{\omega_{\cal B}}^2}}\over{{\hbar}^2}} ,\qquad
D = {-}{ { 4 {k_y} m {\omega_{\cal B} }}\over{\hbar}},  \qquad
{\hat{E}} =  {{ 2 m E }\over{{\hbar}^2}} - {( {{k_y}^2} +  {{k_z}^2} )} .
\end{eqnarray}

 The generation of the spectrum associated with a second order differential
equation of the form
\begin{eqnarray}
{{{d^2}{\cal R}}\over{ds}^2} + f(s) {\cal R}  = 0 \label{eq:calR}
\end{eqnarray}
where 
\begin{eqnarray}
f(s)  = {a\over{s^2}} + b s^2  +c
\end{eqnarray}
has been analysed by several authors \cite{wybourne}. To bring equation
(\ref{eq:BE}) to the above form we set
\begin{eqnarray}
x = s^2   , \qquad  u(x) = {s}^{1\over 2} {\cal R } (s)
\end{eqnarray}
to get
\begin{eqnarray}
{{{d^2}{\cal R}}\over{{ds}^2}}  
+ {[ {( {{ 16 C - 3 }\over 4} )} {1\over{s^2}}
+ 4 {\hat E} {s^2} 
+ 4 D ]} {\cal R} = 0  \label{eq:calR1} .
\end{eqnarray}

We indicate in brief the procedure to obtain the eigenvalues.
The Lie algebra of non-compact groups $SO(2,1) $ and $SU(1,1)$
can be realized in terms of a single variable by expressing the generators
\cite{wyb150} 
\begin{eqnarray}
{{\Gamma}_1} = {{{\p}^2}\over{{\p s}^2}} + {{\alpha}\over{s^2}}
               + {{s^2}\over{16}} \\
{{\Gamma}_2} = - {i\over 2 } {(s {{\p}\over{\p s}} + {1\over 2} ) } \\
{{\Gamma}_3} =  {{{\p}^2}\over{{\p s}^2}} + {{\alpha}\over{s^2}}
               - {{s^2}\over{16}}
\end{eqnarray}
so that the $\Gamma$'s satisfy the standard algebra
\begin{eqnarray}
{[{{\Gamma}_1} ,{{\Gamma}_2} ]} = - i {{\Gamma}_3 } , \qquad
{[{{\Gamma}_2} ,{{\Gamma}_3} ]} =  i {{\Gamma}_1 } ,   \qquad
{[{{\Gamma}_3} ,{{\Gamma}_1} ]} =  i {{\Gamma}_2 } .
\end{eqnarray}
The existence of the Casimir invariant for $su(1,1)$ 
\begin{eqnarray}
{{\Gamma}^2} = {{{\Gamma}_3}^2} -  {{{\Gamma}_1}^2} - {{{\Gamma}_2}^2}
\end{eqnarray}
is exploited to obtain the explicit form of $ {{\Gamma}_i}$'s.
The second order differential operator in equation (\ref{eq:calR})
in terms of the $su(1,1)$ generators is now
\begin{eqnarray}
 {{{\p}^2}\over{{\p s}^2}} +       {a\over{s^2}} + b s^2  +c 
 =  {({1\over 2} + 8 b )} {{\Gamma}_1} +{({1\over 2} - 8 b )} {{\Gamma}_3} 
+ c
\end{eqnarray}
and (\ref{eq:calR} ) becomes 
\begin{eqnarray}
 {[{({1\over 2} + 8 b )} {{\Gamma}_1} +{({1\over 2} - 8 b )} {{\Gamma}_3 }
+ c  ]}{\cal R} = 0  .
\end{eqnarray}
Next a transformation involving $ {e}^{-i \theta {\Gamma}_2 } $ can be
 performed
on $\cal R $ and the ${\Gamma}$'s . A choice of $ \theta $ such that
\begin{eqnarray}
{\tanh{\theta}} = - {  {{1\over 2} + 8 b}\over {{1\over 2} - 8b }}
\end{eqnarray}
will diagonalize the compact operator ${\Gamma}_3 $ and the discrete
 eigenvalues may be obtained. The arguments of the standard
 representation theory 
then leads to the result,
\begin{eqnarray}
4n + 2 + {\sqrt{1 - 4 a } } = { c\over {\sqrt{ - b}}} ,\qquad
 n = 0, 1, 2, \dots
\end{eqnarray}
Substitution of the corresponding values from equation (\ref{eq:calR1})
\begin{eqnarray}
 a =  {( {{ 16 C - 3 }\over 4} )} ,\qquad  b = 4 \hat E ,\qquad
 c = 4 D  \label{eq:a}
\end{eqnarray}
gives
\begin{eqnarray}
{\hat E} = {- } { { 16 m^2 {{\omega}_{\cal B}}^2  {k_y}^2  }\over{
{ \hbar^2   { [ ( 2n + 1 ) +
      {( 1  + { { 16 m^2 {{\omega}_{\cal B}}^2}\over \hbar^2 }}  )}^{1\over 2 }
      ]}^2 }   }  .
\end{eqnarray}

Having obtained the energy eigenvalues, an analysis in the context of 
 two dimensional confined systems on the lines of de Haas - van Alphen
effect and quantum Hall effects would be of much interest from the physics 
point of view.

\section{Group Analysis}
     
Since symmetries and symmetry breakings play such a fundamental role
in physics, we are interested in finding the symmetry generators in
the presence and in the absence of the fields.

We try to find the symmetries of the classical electrodynamics
for a charged particle moving in the presence of a magnetic field.

     The equation of motion of a charged particle with charge $e$ and
 mass $m$
 in a magnetic field $\bf{B}$ is the Lorentz force equation

\begin{eqnarray}
    \ddot{q^a}  &=&  {e\over m} {\varepsilon}^{abc}{\dot{q^b}} {B^c}
    \label{eq: a} 
\end{eqnarray}
where $ a = 1, 2, $ and $ 3$, and $ q^1 = x $, $q^2 = y$, and $q^3 = z $, \\
with a dot representing differentiation with respect
 to time. $\bf{B} $ in our case is given by 
\begin{eqnarray}
 {\bf{B}} =  \{ B_x , B_y , B_z \}  =   \{ 0 , 0, {-} {{\cal B}\over{x^2}}  \} 
\end{eqnarray}
 It is convenient to write the set of coupled equations in the following form
to perform the group analysis.
\begin{eqnarray}
{\ddot{q^{a}}} &=&  {e\over m} {{\omega}^{a}} {({{q^{i}}},{\dot{q^{i}}},t )}
\end{eqnarray}
 where $ {a,i = 1,2,}$
and  $3$, and $ {{\omega}^{a}} $ is equal to the right hand side of equation
 ( \ref{eq: a}) .
Henceforth we follow the notation and method of Stephani{\cite{stephani}}
to find the symmetry generators.

These set of equations can be analysed by means of one parameter groups
by infinitesimal transformations. We demand the equation to be invariant under
infinitesimal changes of the explicit variable ${t}$, as well as simultaneous
infinitesimal changes of the dependent functions ${q^{a}}$ in the following
way,

\begin{eqnarray}
t \rightarrow t_1 &=&  t + \epsilon {\xi} {(t,q^1 ,q^2 ,q^3)}
 + O {({\epsilon}^2)},
\nonumber\\
{{q}^{a}} {\rightarrow} q^a_1 &=& {{q}^{a}} + {\epsilon}{{\eta}^{a} {(t,q^i)}}
 + {O {({\epsilon}^2)} }   \label{eq: b} .
\end{eqnarray}

Under $t \rightarrow t_1$ and $q^a \rightarrow q^a_1$, the
equation changes to,
\begin{eqnarray}
{\ddot q}^a_1 &=&  \omega^a_1 (t_1, q^i_1, \dot q^i_1)
\end{eqnarray}
where we have set $ e\over m$ as unity.
To illustrate the procedure consider the simple case in one space dimension.
We express the above equation in terms of ${ t} $ and ${q}$ by using the
transformation 
 ( \ref{eq: b}) . Then the invariance condition implies that an expression
containing various partial derivatives of ${\xi}$ and ${\eta}$  is obtained
which equates to zero. For example, we get
\begin{eqnarray}
{ \frac{d{q_1}}{d{t_1}}} &=&
 { \frac{ dq + {\epsilon}{( {\frac{\p{\eta}}{\p{t}}} dt
+ \frac{\p{\eta}}{\p{q}}  dq )} }{ dt +
 {\epsilon}{(\frac{\p{\xi}}{\p{t}} dt
+ \frac{\p{\xi}}{\p{q}} dq )} }}   +  O ({\epsilon}^2) .
\end{eqnarray}
and now relate the left hand side with $\frac{dq}{dt} $ by using binomial
theorem for the denominator to obtain
\begin{eqnarray}
\frac{d{q}_{1}}{d{t_1}}  &=&  \frac{dq}{dt}
 + \epsilon {( \frac{\p{\eta}}{\p{q}}
- \frac{\p{\xi}}{\p t})}{ \frac{dq}{dt}} -  {\frac{\p\xi}{\p q}}
{(\frac{\p q}{\p t})}^2  + O({\epsilon}^2)   . \label{eq: c}
\end{eqnarray}
A similar procedure is followed to express
$\frac{{d}^2 {q_1}}{{{dt}_1}^2} $
likewise. By substituting equations 
 ( \ref{eq: b})  - (\ref{eq: c}) for a given explicit expression
for $ {{\omega}_1} $ and remembering that  $ {\frac{d^2 q}{dt^2}  - \omega} $
is zero, we obtain the desired partial differential equation whose solution
would determine $ {\xi{(t,q)}} $ and $ {\eta{(t,q)}} $. In our case, of course,
we have to find $ {\xi{(t,q^1 ,q^2 ,q^3 )}} $ and
$ {{\eta}^a}{(t,q^1 ,q^2 ,q^3 )} $'s.

To relate these to the generators of the infinitesimal transformations
we write
\begin{eqnarray}
t_1 {(t, q^i ; \epsilon )} =  t +
 \epsilon {\xi (t , q^i )} +  \cdots
= t + \epsilon{ \bf{X}} t + \cdots  \\
{{{q^a}_1} {({{t}{,}{q^i} {;}{\epsilon}}) }} = {{q^a}  +{\epsilon}
 {{\eta} {({{t}{,}{q^i}} )}} + {\cdots}}
= {{q^a} + {\epsilon} {\bf{X}}{ q^a } + {\cdots}}
\end{eqnarray}
where the functions $\xi$ and ${\eta}^a $ are defined by
\begin{eqnarray}
{\xi (t,q^i ) } &=& 
 {{ \frac{\p{t_1}}{\p{\epsilon}}}{{\mid}_{\epsilon = 0}}} ,\\ 
{{{\eta}^a}(t,q^i )} &=&
 {{\frac{\p{{q^a}_1}}{\p{\epsilon}}}{{\mid}_{\epsilon =0}}}
\end{eqnarray}
and the operator $\bf{X}$ is given by
\begin{eqnarray}
{\bf{X}} &=& {{\xi (t,q^i )} {\frac{\p}{\p{t}}}  
  + {{\eta}^a (t,q^i )}} {\frac{\p}
{\p{q^a}}}.
\end{eqnarray}
Following Stephani \cite{stephani}, we will consider the equation having the
symmetry generated by $\bf{X} $ and its extension
\begin{eqnarray}
{\dot{\bf{X}}} &=& \xi {\frac{\p}{\p{ t}}}
 + {{\eta}^a} {\frac{\p}{{\p}{q^a}}} + 
{\dot{\eta^a}}{\frac{\p}{{\p{\dot{q^a}}}}}
\end{eqnarray}
and the symmetry condition determines ${\dot{\eta}^a} $.
 The symmetry condition
is given by
\begin{eqnarray}
{\xi} {{{\omega}^a}_{,t}} + {\eta}^b {{\omega}^a}_{,b} + {( {{\eta}^b}_{,t}
+ {\dot{{q}^{c}}} {{\eta}^b}_{,c} - {\dot{q^b}} {{\xi}_{,t}}
-  {\dot{q^b}}{\dot{q^c}} {{\xi}_{,c}} )}
{ \frac{\p{{\omega}^a}}{\p{\dot{q^b}}}}\nonumber\\
+ 2 {{\omega}^a} {( {{\xi}_{,t} +  {\dot{q^b}}}{{\xi}_{,b}} )}
+ {\omega^b}{(\dot{q^a}{{\xi}_{,b}} - {{\eta}^a}_{,b}) }
+ {\dot{q^a}} {\dot{q^b}} {\dot{q^c}} {\xi}_{,bc}\nonumber\\
+ 2 {\dot{q^a}}{\dot{q^c}} {{\xi}_{,tc}}  - 
  {\dot{q^c}} {\dot{q^b}} {{\eta}^a}_{,bc}
+ {\dot{q^a}}{{\xi}_{,tt}} - 2 {\dot{q^b}} {{\eta}^a}_{,tb}
- {{{\eta}^a}_{,tt}}  &=& 0  \label{eq: wa}
\end{eqnarray}
where $ f_{,t} = {\frac{{\p}f}{\p{t}}} $ and $ f_{,c} =
 {\frac{{\p}f}{\p{q^c}}} $.
By herding together coefficients of the terms with cubic, quartic, and linear
in $ {\dot{q^a}} $ ,
and the ones  independent of $ \dot{q^a} $ separately,
 and equating each of these
to zero we obtain an over determined set of partial differential equations
and solve for $ {\xi} $ and $ {{\eta}^{a}} $.
  The condition (\ref{eq: wa}) for $a = 1 $, and $2$ gives 
\begin{eqnarray}
 \xi = \lambda , \qquad
 \eta^1 = 0 ,  \qquad  \eta^2 = \sigma ,       
\end{eqnarray}
and for $a = 3$ we obtain
\begin{eqnarray} 
\eta^3 =  \rho + q^3 
\end{eqnarray}
where $ \lambda $, $\sigma$, and $\rho $ are constants.
This gives rise to the vector fields
\begin{eqnarray}
 {\bf X }_{\xi} = \lambda {{\p}\over{\p t}} ,\qquad 
 {\bf X }_{{\eta}^2}  = \sigma  {{\p}\over{\p {q^2}}} , \qquad
 {\bf X }_{3}  = \rho  {{\p}\over{\p {q^3}}}, \qquad
  {\bf X }_{{\eta}^3} = q^3  {{\p}\over{\p {q^3}}}, 
\end{eqnarray}
that forms a solvable Lie algebra.

This may be compared with the results for the cases where
 no magnetic field is present and for a magnetic field proportional to
$ q^a$ . In the case of no magnetic field present, the equations
 trivially decouple.
Each equation
\begin{eqnarray}
  \ddot{q^a}  = 0
\end{eqnarray}
 has the eight parameter symmetry generator of the general projective
transformation 
\begin{eqnarray}
{\bf X } = [ {a_1} + {a_2} t + {a_3} {q^a} + {a_4} t {q^a} + {a_5} {t^2} ]
 {{\p}\over{\p t}}  \nonumber  \\
 + [ {a_6} + {a_7} t + {a_8} {q^a} + {a_5}t {q^a} + {a_4} {{(q^a)}^{2}}]
{{\p}\over{\p{q^a}}}  .
\end{eqnarray}

 For $  B^a  = q^a $ the five  generators are \cite{mah} 
 
\begin{eqnarray}
{{\bf{X}}_{q^a}} &=&
 {{{{\varepsilon}_{a}}^{k}}_{b}} {q^b} {\frac{\p}{\p{q^k}}}\\
{{\bf{X}}_{\xi}}  &=& {\frac{\p}{\p{t}}}   \\
{{\bf{X}}_5}  &=& {t}{\frac{\p}{\p{t}}} - {q^a} {\frac{\p}{\p{q^a}}} .
\end{eqnarray}

In order to find the Lie point symmetries of the corresponding
quantum case we go back to the Schr\"odinger equation (\ref{eq:BE}).
We write this equation as
\begin{eqnarray}
u'' = \omega (x,u,u') \label{eq:our}
\end{eqnarray}
where
\begin{eqnarray}
{\omega} (x,u,u') = - { ( {C\over{x^2}} + {D\over{x}} + \hat{E})} u(x) .
\end{eqnarray}
The infinitesimal generator of the symmetry under which the differential 
equation does not change is given by the vector field
\begin{eqnarray}
{\bf{X} } = {\xi }(x,u) {{\p}\over{\p x} } + {{\eta}(x,u) {{\p}\over{\p u}}}
\end{eqnarray}
and for a second order differential equation, $ \xi $ and $\eta$ are to
be determined from
\begin{eqnarray}
\omega ( {\eta}_{,u} - 2 {\xi}_{,x} -3 u' {\xi}_{,u} )
 - {\omega}_{,x} \xi - {\omega}_{,u} {\eta}
- {\omega}_{,{u'}} {[ {\eta}_{,x} + u' ( {\eta}_{,u} -{\xi}_{,x} )
 - {u'}^2 {\xi}_{,u} ]}     \nonumber     \\
+ {\eta}_{,xx} + u' ( 2 {\eta}_{,xu} - {\xi}_{,xx} )
+ {u'}^2 ( {\eta}_{,uu} - 2 {\xi}_{,xu} ) - {u'}^3 {\xi}_{,uu}
= 0  \label{eq:gen}
\end{eqnarray}
where a prime denotes differentiation with respect to $x$,
 the partial derivative of a function by a comma followed by
 the variable with respect to which the derivation has been performed.

The symmetry condition (\ref{eq:gen}) for  our equation (\ref{eq:our}) is
\begin{eqnarray} 
 - [{ ( {C\over{x^2}} + {D\over{x}} + \hat{E})} u(x) ]
 ({\eta}_{,u} - {\xi}_{,x} )  + [ {2C\over{x^3}} + {D\over{x^2}} ] (u \xi )
+ { ( {C\over{x^2}} + {D\over{x}} + \hat{E})} {\eta}
+ {\eta}_{,xx}  \nonumber  \\
- u' [ 3 ({C\over{x^2}} + {D\over{x}} + \hat{E}) {\xi}_{,u} + 2 {\eta}_{,xu}
- {\xi}_{,xx } ] 
+ {u'}^2 ( {\eta}_{,uu} - 2 {\xi}_{,xu} ) 
- {u'}^3 {\xi}_{,uu} = 0  .
\end{eqnarray}
Equating to zero the coefficients of ${u'}^3 $ and ${u'}^2 $
we get
\begin{eqnarray}
{\xi}_{,uu} =0 , \qquad {\eta}_{,uu} = 2 {\xi}_{,xu}
\end{eqnarray}
which are satisfiesd for
\begin{eqnarray}
{\xi} = u \alpha (x) + \beta (x) , \qquad  \eta  = u^2  {\alpha}' (x) +
 u {\gamma}(x) + \delta (x)  .
\end{eqnarray}
Using these and equating to zero the coefficient of $u'$ one obtains
\begin{eqnarray}
3 u [ {\alpha}'' (x) +   ( {C\over{x^2}} + {D\over{x}} + \hat{E}) {\alpha} (x)
] + 2 {\gamma}' (x) - 2 {\beta}'' (x) = 0  .
\end{eqnarray}
This shows that either 
\begin{eqnarray}
{\alpha} = 0
\end{eqnarray}
or
$\alpha (x) $ satisfies the same equation as $u(x)$ does and
\begin{eqnarray}
2 {\gamma}' (x) = {\beta}'' (x)  .
\end{eqnarray}
 This integrates to
\begin{eqnarray}
\gamma (x) =  {{{\beta}' (x)}\over 2} + \kappa
\end{eqnarray}
where $\kappa $ is a constant. The rest of the equation, after using the
fact that $\alpha (x) $ satisfies the same equation as $u (x) $ does ,
 boils down
to
\begin{eqnarray}
{\beta}''' (x) + 4  { ( {C\over{x^2}} + {D\over{x}} + \hat{E})} {\beta}' (x)
- ({{4C}\over {x^3}} + {{2D}\over{x^2}} ) {\beta} (x) = 0 \label{eq:beta}
\end{eqnarray}
with $\delta (x) $ also obeying the same equation as $u (x) $ does.
The simplest nontrivial solution to this equation is
\begin{eqnarray}
\beta (x) = {p\over x } + q
\end{eqnarray}
where $p$ and $q$ are constants and  $C$ ,$D$ , and $\hat E $ must satisfy
\begin{eqnarray}
q = 2 p D , \qquad  C = - {3\over 4}, \qquad  D^2  = - \hat E  .
\end{eqnarray}
One sees that these conditions are in principle possible to be satisfied.
It is further worth noting that the condition $ D^2 = - \hat E $ is satisfied
for $ C = - {3\over 4} $ if we take the square root to be only negative  while
evaluating $ \sqrt{ 1 - 4 a } $ and for $ n = 1 $ . Also note that the symmetry
is present for a particular value of the energy. It may be  recalled that
$ n $ must be zero or a positive integer for our analysis if any meaningful 
eigenvalue is to be obtained. The above result indicates that only for
$ n = 1 $ there is the symmetry corresponding to  $ \beta = {p\over{x^2}}
+ q $. So the original equation ({\ref{eq:our}}) has the symmetry 
represented by the vector field 
\begin{eqnarray}
{\bf X} = {[ u {\alpha (x) } + {p\over x} + 2 p D ]} {{\p }\over{\p x}}
          + { [ u^2 {\alpha}' (x) - u {p\over{x^2} } + u \kappa 
          + {\delta} (x) ] } {{\p}\over{\p u}} .  \label{eq:al}
\end{eqnarray}
Here $u$, $\alpha$, and $\delta$ are related to the hypergeometric
functions $_1F_1 $  which are solutions to generalised Kummer type 
of equations mentioned in section 2.
 We  immediately observe that the generators do not close under the Lie
product.

We also find  that 
\begin{eqnarray}
\beta (x) = { {g_2}\over{x^2}} + {{g_1}\over{x} } + g_0
\end{eqnarray}
will satisfy the equation (\ref{eq:beta})  and everything will be consistent
if 
\begin{eqnarray}
g_0 = {-} {{2 \hat E }\over{D}} g_1  ,{\qquad } g_1 = 2D g_2 ,{\qquad }
 C = -2  .
\end{eqnarray}
This would imply 
\begin{eqnarray}
\hat E  = - {{d^2}\over 4}  .
\end{eqnarray}
This can only be obtained if we again take the negative root of
 $\sqrt { 1 - 4 a }$ and $n = 2$.Hence there is again an enhancement
of symmetry at another possible eigenvalue and for another value of n.
The vector-field for this case is
\begin{eqnarray}
{\bf X} = {[ u {\alpha (x) } + ( {1\over{4 {\hat E}  x^2}} -
 {2 {\hat E}\over{ Dx} }  ){g_0} ]} {{\p }\over{\p x}} \nonumber  \\
         	 + { [ u^2 {\alpha}' (x) - u ( \{ {1\over{2{\hat E}x^3} } + 
 - {{2 {\hat E}\over{Dx^2}}} \}{g_0}      -    \kappa )  
          + {\delta} (x) ] } {{\p}\over{\p u}}   \label{eq:al2}
\end{eqnarray}

 and again
all Lie products do not close.

Taking 
\begin{eqnarray}
\beta (x) = { {g_3}\over{x^3}} + { {g_2}\over{x^2}} + {{g_1}\over{x} } + g_0
\end{eqnarray}
the relation between $g_l$'s with $l = 0, 1, 2,$ and $3$ , becomes
\begin{eqnarray}
 {g_3} = { 3\over{2D}} {g_2}, \qquad
 {g_2} = { 12 D \over{9 {\hat E} + 5 {D^2}}}{g_1} ,  \qquad
 {g_1} =  {5\over{[ {32 D \hat E \over{ 9 \hat E + 5 D^2 }} + 2D ]}}{g_0},
 \qquad
 { C} = {- 15\over{4}} .
\end{eqnarray}
All the previous considerations apply in this case with $n = 3 $.
The vector field can be easily determined.

Hence, in all above cases  , we find that the symmetry gets enhanced
at particular eigenvalues.
 
 We expect that similar analysis can be performed by including the 
higher negative powers
of $x$  , such as,
\begin{eqnarray}
 \beta (x) = \sum_{n} { {g_{n}} x^{-n} } ,  \qquad n = 0, 1, 2, \dots  . 
\end{eqnarray}
 as a solution of (\ref{eq:beta}).

In the absence of any magnetic field, equation (\ref{eq:our}) reduces to
\begin{eqnarray}
u'' + u = 0
\end{eqnarray}
where we have scaled $u$ so that $\hat E$ becomes equal to unity.
For comparison, the vector field in this case is 
\begin{eqnarray}
{\bf {X}} = {[{ a_1} u \sin{( x + {a_2} )} + {a_7} ( \sin{ 2x + {a_8}} )
 + {a_6} ]} 
{{\p}\over{\p x}}  \nonumber  \\
 + { [ {a_1} {u^2} \cos{( x + {a_2}} ) + u \{ {a_7}  \sin{ (2 x + {a_8}} )
 + {a_5} \}
 + {a_3} \sin{ ( x + {a_4}} ) ]} {{\p}\over{\p u}}  .
\end{eqnarray}
However, the algebra we obtain, (\ref{eq:al}) , looks more interesting.

\section{Conclusion}
        
The physical system of a charged particle under the influence
of a  constant magnetic field, which was considered  early
after the foundation of quantum mechanics,   has recent  
ramifications in quantum Hall effects
and in the gauge theories on noncommutative spaces in 
connection with the quantization
of D-branes\cite{bigatti}. One would expect that the study of 
the system under a
 space dependent
magnetic field may  result in  more pronounced and interesting behaviour. 
We consider a simple situation where only the $y$ - component
of the vector potential exists in the form $B_z$ proportional
 to $1\over{x^2} $. Using the method of group analysis for the 
classical equation of motion,
we find that the symmetry is drastically reduced in this case .
Such analysis has also been carried out for magnetic field proportional to
coordinates, $ B^a \propto q^a $. More symmetric and interesting results
were obtained \cite{mah} in that case.

In the present case a simple group theoretic calculation gives the eigenvalues
for the quantum mechanical problem. Further, the group analysis of
 the differential equation shows how the symmetries are enhanced 
at certain values of energy
and correspondingly  terms get added to the vectorfield
characterising the symmetry. For the lowest values of $n$ we
 also obtain vector fields that do not close
under Lie product. The structure of the vector fields reminds us of
the algebras of Kac - Moody - Virasoro  type \cite{god}.

\noindent{\bf Acknowledgement}
 
I would like to thank Professor David S. Jerison and Professor David
A. Vogan for kind invitation to visit Massachusetts Institute of Technology,
where this work was carried out. I am also grateful to Professor D. A. Vogan 
 for  discussions. It is a pleasure to thank Professor B. Boghosian
and Professor P. Love  for helpful comments.


\end{document}